\documentclass[twocolumn,showpacs,preprintnumbers,amsmath,amssymb,prb,superscriptaddress]{revtex4}

\usepackage{amsmath}
\usepackage{graphicx}
\usepackage{dcolumn}
\usepackage{bm} 

\newcommand{\degree}{{$^\circ$}}
\newcommand{\monolayer}{Sr$_2$RuO$_4$}
\newcommand{\bilayer}{Sr$_3$Ru$_2$O$_7$}
\newcommand{\inflayer}{SrRuO$_3$}

\begin{document}


\title{\boldmath Orbital Properties of Sr$_3$Ru$_2$O$_7$ and
Related Ruthenates Probed by $^{17}$O-NMR}

\author{K.~Kitagawa}
\affiliation{Department of Physics, Graduate School of Science, Kyoto
University, Kyoto 606-8502, Japan} 
\email{kitagawa@scphys.kyoto-u.ac.jp}

\author{K.~Ishida}
\affiliation{Department of Physics, Graduate School of Science, Kyoto 
University, Kyoto 606-8502, Japan}
\affiliation{Kyoto University International Innovation Center, Kyoto 606-8501, Japan}

\author{R.~S.~Perry}
\affiliation{Department of Physics, Graduate School of Science, Kyoto University, Kyoto 606-8502, Japan}
\affiliation{Kyoto University International Innovation Center, Kyoto 606-8501, Japan}
\affiliation{School of Physics and Astronomy, University of St.\,Andrews, Fife KY16 9SS, Scotland}

\author{H.~Murakawa}
\affiliation{Department of Physics, Graduate School of Science, Kyoto 
University, Kyoto 606-8502, Japan}

\author{K.~Yoshimura}
\affiliation{Department of Chemistry, Graduate School of Science, Kyoto 
University, Kyoto 606-8502, Japan}

\author{Y.~Maeno}
\affiliation{Department of Physics, Graduate School of Science, Kyoto University, Kyoto 606-8502, Japan}
\affiliation{Kyoto University International Innovation Center, Kyoto 606-8501,
Japan}

\date{\today}

\begin{abstract}
We report a site-separated $^{17}$O-NMR study of the layered perovskite ruthenate
Sr$_3$Ru$_2$O$_7$, which exhibits nearly two-dimensional transport properties
and 
itinerant metamagnetism at low temperatures. The local hole occupancies and the
spin
densities in the oxygen $2p$ orbitals are obtained by means of tight-binding
analyses of electric field gradients and anisotropic Knight shifts.
These quantities are compared with two other layered perovskite ruthenates: the 
two-dimensional paramagnet Sr$_2$RuO$_4$ and the three-dimensional ferromagnet 
SrRuO$_3$. The hole occupancies at the oxygen sites are
very
large, about one hole per ruthenium atom. This is due to the strong covalent
character
of the Ru-O bonding in this compound. The magnitude of the hole occupancy might be
related to the rotation or tilt of the RuO$_6$
octahedra. The spin densities at the oxygen sites are also large,
20-40\% of the bulk susceptibilities, but in contrast to the hole occupancies,
the spin densities strongly depend on the dimensionality. This result suggests that the 
density-of-states at the oxygen sites plays
an essential role for the understanding of the complex magnetism found in the
layered perovskite ruthenates.
\end{abstract}


\maketitle

\section{Introduction}
The Ruddlesden-Popper (R-P) series of layered perovskite ruthenates has
attracted great interest because of their wide variety of phenomena. The
strontium ruthenates Sr$_{n+1}$Ru$_n$O$_{3n+1}$ are metallic members of
the R-P perovskite ruthenates featuring various kinds of itinerant magnetism and
magnetic fluctuations.
The two-dimensional member ($n=1$) \monolayer\ is paramagnetic and exhibits
unconventional superconductivity attributable to $p$-wave pairing,
\cite{214maeno,214review,214ishida} originating from electron correlations in the
Ru $4d_{xy}$ band. At the same time, it possesses strong
incommensurate antiferromagnetic fluctuations\cite{214SidisINS} due to
nesting of the $4d_{yz,zx}$ bands. The three-dimensional end member ($n=\infty$)
\inflayer\ is an
itinerant ferromagnet with a Curie temperature of 
160~K.\cite{113}

The intermediate dimensional, bilayer member
($n=2$) \bilayer\ does not exhibit magnetic order at zero magnetic field and
ambient pressure. Nevertheless, \bilayer\ is considered to be very close to a
ferromagnetic instability: (a) the uniform susceptibility has a peak around
16~K, (b)
the Wilson ratio is quite large $\sim 10$, and (c) along $c$-axis applying small
uniaxial
pressure of about 0.1~GPa pushes \bilayer\ into a ferromagnetic phase below
80~K.\cite{SIkeda, SIkedaPressure} At ambient
pressure, field-induced metamagnetic transitions are
observed.\cite{PerryPRL,PerryJPSJ} The investigation of those metamagnetic
transitions
leads to a phase diagram with a line of first-order phase transitions terminating
at a critical end point. Furthermore, the critical end temperature
$T^*$ is extremely low: $T^*=1.2$~K for $H \perp c$ and $T^*$ vanishes for $H
\parallel c$.\cite{GrigeraPRB} In the latter case ($H \parallel c$), the
existence of a metamagnetic quantum critical point (MMQCP) was proposed.\cite{MillisMMQCP,GrigeraQCP} 
In our previous $^{17}$O-NMR study with $H \parallel c$, we reported the measurement of the
nuclear spin-lattice relaxation rate $1/T_1$, which probes the spin
fluctuations, and concluded that two-dimensional ferromagnetic character is dominant at high
temperatures. In
contrast, non-ferromagnetic (mostly incommensurate antiferromagnetic)
fluctuations diverge for $T\rightarrow 0$~K near the
MMQCP.\cite{KitagawaPRL} 

 Such competitions between itinerant ferromagnetism and incommensurate antiferromagnetism in
 Sr$_{n+1}$Ru$_n$O$_{3n+1}$ are a result of the existence of narrow bands and a
 high density-of-states (DOS) at the Fermi level $E_{\rm F}$ created by
 ruthenium $4d$ and oxygen $2p$ orbitals. According to a band-structure
 calculation for \monolayer,\cite{Oguchi214} $4d$-$2p$ antibonding bands at the Fermi
 level create 18\% of the DOS in the oxygen $2p$ orbitals and possibly play
 a crucial role for the electronic and/or structural instabilities.
 The large DOS at the oxygen sites is unlikely for $3d$ transition-metal
 oxides because the
 covalency between the $d$ and $2p$ orbitals is usually rather weak. Our concerns
 are: (i) How does the hole occupancy or DOS in the oxygen $2p$ orbitals
 affect the occurrence of itinerant ferromagnetism or antiferromagnetism and
 (ii) what is the relationship between dimensionality and magnetism.
 
 In this article, we report a detailed site-separated $^{17}$O-NMR study of
 \bilayer. In Sec.~\ref{sec:efg}, we estimate the
hole occupancies in the $2p$ orbitals from the NMR quadrupole splittings at the
oxygen sites. Since the oxygen $2p$ orbitals feature a magnetic-dipole
type of hyperfine interaction, the anisotropy of the Knight shifts needs to be considered. In Sec.~\ref{sec:shift} we obtain 
 the spin susceptibility
of each orbital at all oxygen sites from the Knight shifts. For comparison, we also analyze the
related ruthenates \monolayer\ and \inflayer\ applying the same method.

\section{Experiment}
In our NMR experiment we used singe crystals of \bilayer\  which were grown in
a floating-zone image
furnace.\cite{PerryGrowth} We exchanged natural oxygen $^{16}$O ($I=0$) for
$^{17}$O ($I = 5/2$). The substitution was performed
in a silica-glass-tube furnace, where the samples were kept for one week at
1050\degree C in a 70\% concentrated $^{17}$O$_2$ atmosphere. 
Although the gain of mass, estimated to be 0.9\%, could not be observed,
the measured $^{17}$O-NMR signals assure that the isotope substitution process
was successful at least over the penetration depth accessible in our
experiment, {\it i.\,e.} several micro meters. This is sufficient to ignore
surface related effects. After the annealing process, we confirmed the residual
in-plane resistivity,
to be $0.8$~$\mu\Omega\cdot$cm. This is enough clean to observe the intrinsic
behavior of this system.\cite{PerrySdH}

Great care was taken to keep the orientation of the sample throughout the field-dependent
measurements. For this purpose, complex rf
 components were introduced outside the cryostat. This ensured that the
 measurements for all
 desired field strengths, corresponding to frequencies of 10-80~MHz, could be
 carried out using only one probing coil.
 A homemade wide-band pre-amplifier designed with
 high-electron-mobility transistors was immersed in liquid nitrogen to obtain an
 excellent noise figure of 0.5~dB. Furthermore it had a short dead-time of
 less than 3~$\mu$s
 after saturation. With this setup it was possible to overcome the issues of the
 weak intensities of signals and shortness of NMR decoherence times.

\section{Results and Discussions}
\subsection{NMR spectra and site-assignment}
\begin{figure}[tbp] \centering
\includegraphics[width=0.8\linewidth]{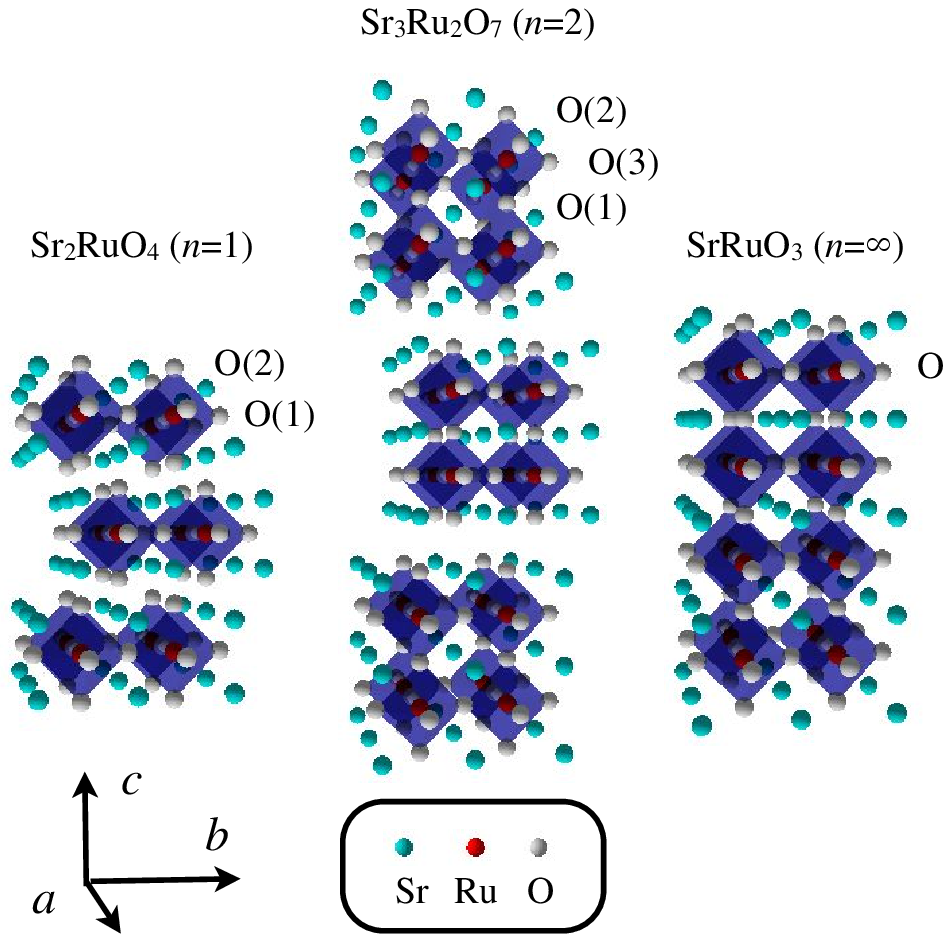} 
\caption{\label{fig:ruthenates}
(color online). Crystal structure of the Ruddlesden-Popper series
Sr$_{n+1}$Ru$_n$O$_{3n+1}$ for 
\monolayer\ ($n=1$, left), \bilayer\ ($n=2$, middle), and \inflayer\
($n=\infty$, right). The axes $a$ and $b$ are taken along a Ru-O-Ru bonding. For
\bilayer\ and \inflayer, symmetries are reduced because of a rotation and a
tilt of the RuO$_6$ octahedra, respectively.\cite{Shaked327,Jones113} For
simplicity, these effects are not shown. 
There are apical O(2) and in-plane O(1) sites in \monolayer. \bilayer\ has two
kinds of apical sites. For \bilayer, we label the
inner-apical, outer-apical, and in-plane sites O(1), O(2), and O(3) sites, 
respectively.} 
\end{figure}

$^{17}$O-NMR lines can be described using anisotropic
Knight shifts and the quadrupole interaction perturbation of the
electric field gradients (EFG). For
each site, the resonance field $H_{\rm res}$ corresponding to the $m
\leftrightarrow m-1$ resonance can be approximately
 written as\cite{MetallicShifts}\textsuperscript{,}\footnote{Equation~\eqref{eq:res} only contains the first order term with regard
to the perturbations from EFG. For quantitative analyses, the
second order perturbations must be included.
}
\begin{align}\label{eq:res}
&\frac{H_0 - H_{\rm res} }{H_{\rm res} }= {^{17}\!K^{\rm iso}}\nonumber\\
 &+ \left\{\frac{1}{2}\frac{^{17}\nu_{\rm
Q}\left(m-\frac{1}{2}\right)}{^{17}\gamma H_{\rm res} } +
\frac{1}{2}{^{17}\!K^{\rm ax}} \right\}(3\cos^2\theta - 1) \nonumber\\
&-\frac{1}{2}\left\{^{17}\!K^{\rm aniso,\perp}
+\left(m-\frac{1}{2}\right)\frac{^{17}\nu_{\rm Q}}{^{17}\gamma H_{\rm res}}
\eta\right\}\sin^2\theta\cos 2\phi, 
\end{align}
where ($\theta$, $\phi$) are Euler's coordinates with respect to the
largest principal axis
of the diagonalized EFG, and $^{17}\gamma$ ($=5.7719$~MHz/T)
is the gyromagnetic ratio of an $^{17}$O nucleus. $H_0$ is the resonance field
of a free $^{17}$O atom at the resonance frequency
$^{17}\gamma H_0$. $^{17}\!K^{\rm iso}$ is the isotropic Knight
shift, $^{17}\!K^{\rm ax}$ is the axial anisotropic shift, and
$^{17}\!K^{\rm aniso,\perp}$ is the in-plane anisotropic shift
perpendicular to the principal axis for lower symmetric sites.
$^{17}\nu_{\rm Q}$ is the pure quadrupole frequency and $\eta\,(0 \le \eta \le
1)$ is the asymmetric parameter of the EFG.

\begin{figure}[bp]
\centering
\includegraphics[width=\linewidth]{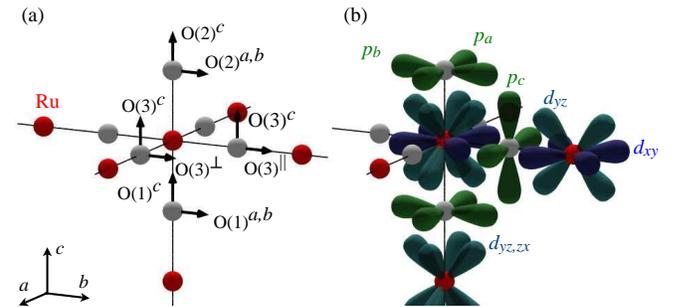} 
\caption{\label{fig:orbitals}
(color online). (a) $^{17}$O-NMR spectral site notations for \bilayer. Arrows
denote the corresponding field directions, $H \parallel c$ and $H \parallel $ [100].
(b) Relevant atomic orbitals at the Fermi surfaces for \bilayer. We treat
interactions of electrons with nuclei by a method of linear combination of atomic
orbitals, namely the tight-binding method.  Ru$^{4+}$ ions realize a low-spin
configuration
and the resulting $t_{2g}$ ($=4d_{xy},4d_{yz},4d_{zx}$) orbitals hybridize with the
oxygen $2p_{\pi}$ ($=2p_x,2p_y,2p_z$) orbitals.}
\end{figure}
\begin{figure}[htbp]
\centering
\includegraphics[width=\linewidth]{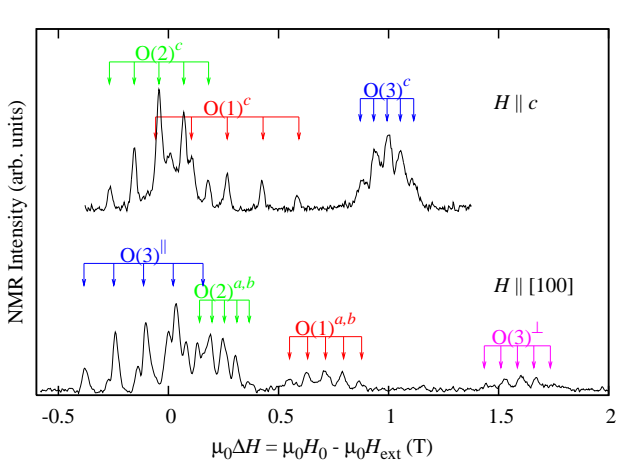}
\caption{\label{fig:spectra}
(color online). Field-swept $^{17}$O-NMR spectra for \bilayer\ with a frequency
of 72.5~MHz at 1.7~K. The
corresponding $^{17}$O-NMR field for a free atom, $\mu_0 H_0$, is 12.56~T.
$H_{\rm ext}$ is the external applied field. Lines from each site must
consist of five resonances due to the perturbations from the electric quadrupole
interaction. The unassigned sites, which exhibit a smaller sample-dependent 
intensity, are attributed to an impurity phase of \monolayer. }
\end{figure}
 Crystallographically, three inequivalent oxygen sites exist in
 the bilayered perovskite structure (Fig.~\ref{fig:ruthenates}). Moreover, when
 a field is applied perpendicular to the $c$ axis, the in-plane site
O(3) splits into two configurations, O(3)$^{\parallel}$ and O(3)$^{\perp}$, as
shown in Fig.~\ref{fig:orbitals}(a).
Figure~\ref{fig:spectra} shows a typical $^{17}$O-NMR spectra for \bilayer. 

For the apical sites (O(1) and O(2) in \bilayer) one can rewrite the parameters
as follows:
\begin{align*}
^{17}\!K^{\rm iso} &= \frac{1}{3}({^{17}\!K_{\rm obs}^{c}} + 2\,{^{17}\!K_{\rm
obs}^{a,b}}),\\
^{17}\!K^{\rm ax} &= {^{17}\!K_{\rm obs}^{c}} - {^{17}\!K^{\rm iso}}\\
&= \frac{2}{3}({^{17}\!K_{\rm obs}^{c}} - {^{17}\!K_{\rm obs}^{a,b}}),\\
^{17}\!K^{\rm aniso,\perp} &= 0,\\
^{17}\nu_{\rm Q} &= {^{17}\!\nu^{c}} = -2\,{^{17}\!\nu^{a,b}},\\
\eta &= 0.
\end{align*}
Here, $^{17}\!K_{\rm obs}^{i}$ is the observed Knight shift for direction $i$.

For in-plane sites (O(3) for \bilayer) the parameters are: 
\begin{align*}
^{17}\!K^{\rm iso} &= \frac{1}{3}({^{17}\!K_{\rm obs}^{\parallel}} +
{^{17}\!K_{\rm obs}^{\perp}} + {^{17}\!K_{\rm obs}^{c}}),\\
^{17}\!K^{\rm ax} &= {^{17}\!K_{\rm obs}^{\parallel}} - {^{17}\!K^{\rm iso}},\\
^{17}\!K^{\rm aniso,\perp} &= {^{17}\!K_{\rm obs}^{\perp}} - {^{17}\!K_{\rm
obs}^{c}},\\
{^{17}\nu_{\rm Q}} &= {^{17}\!\nu^{\parallel}}.
\end{align*}
\begin{figure}[!ht]
\centering
\includegraphics[width=\linewidth]{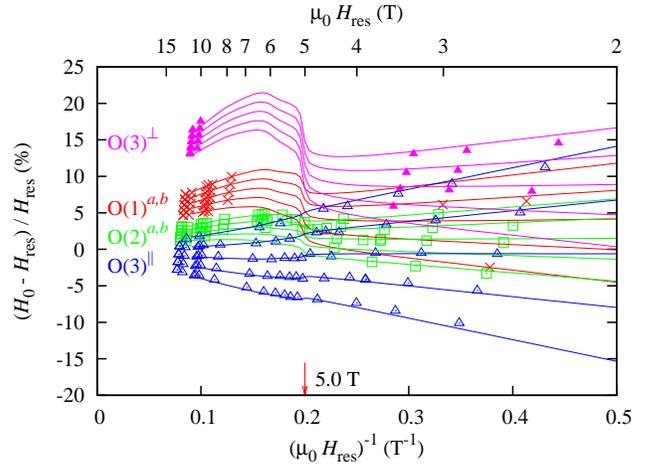}
\caption{\label{fig:kvsh100}
(color online). NMR spectral positions as a function of $H_{\rm res}^{-1}$ for
\bilayer\ with $H \parallel
[100]$. All data points are taken from the field-swept spectra at 1.7~K. The
arrow denotes
the metamagnetic field. At and slightly below the metamagnetic field high
transverse relaxation rates $1/T_2$ prevent observations of signals
from the O(1) and O(3)$^{\perp}$ sites. The plotted lines are based on 
eq.~\eqref{eq:res}. The values of EFGs are listed in
Sec.~\ref{sec:efg}. The Knight shifts are evaluated in Sec.~\ref{sec:shift}.}
\end{figure}
\begin{figure}[ht]
\centering
\includegraphics[width=\linewidth]{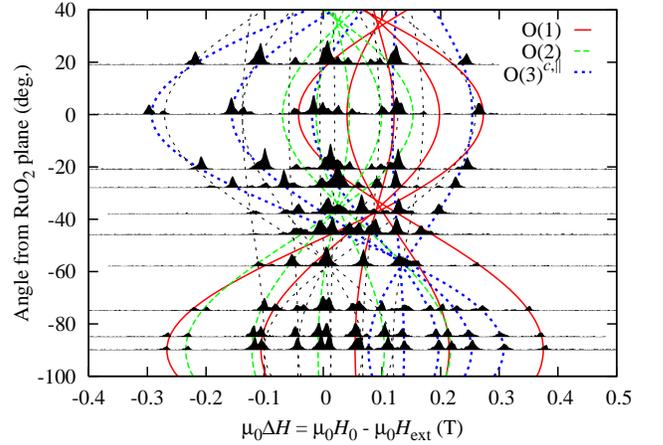}
\caption{\label{fig:variangle}
(color online). Field-swept NMR spectra with various field orientations for \bilayer\ at
1.7~K with a frequency of 21.3~MHz, i.\,e.\ $\mu_0H_0 \sim 3.7$~T.
The orientation is varied from [100] (0~deg. in the plot)
to the $c$ axis ($\pm$90~deg.) by the use of a backlash-free rotator in
 an 8~T split-coil magnet. All curves are drawn in the same scheme as in
 Fig.~\ref{fig:kvsh100}. The O(3)$^{\perp}$ lines are not observable in this
 configuration due to a high transverse relaxation rate $1/T_2$. The narrow
 dotted curves, which are not
 attributable to \bilayer, originate from the \monolayer-O(1) site in the
 impurity phase.}
\end{figure}
From eq.~\eqref{eq:res}, we expect a linear relationship
between $(H_0 - H_{\rm res}) / H_{\rm res}$ and $(H_{\rm res})^{-1}$ as long as the
magnetization $M$ is linear with $H$. Figure~\ref{fig:kvsh100} displays such a behavior
for $H
\parallel [100]$ (see Ref.~\onlinecite{KitagawaPhysicaB} for $H \parallel c$).
The linear relationships are realized much below and above the metamagnetic field
$H_{\rm M}$. The plot is helpful to group spectral lines split up by EFG into
identical sites. Applying different field orientations causes different
responses of the apical and in-plane sites (Fig.~\ref{fig:variangle}). In this way one can distinguish
between them. The difference between the inner-apical O(1) and the
outer-apical O(2) sites appears in the spectral intensities as the corresponding
composition ratio $1:2$, as long as the spectrum is taken under the condition that
the relaxation effects are small. Thus, the three sites in \bilayer\ can be
resolved from the complicated spectra shown in Fig.~\ref{fig:spectra}.

As displayed in Figs.~\ref{fig:spectra} and \ref{fig:variangle}, the
impurity eutectic phase of \monolayer\ sometimes appears in NMR spectra, although
the fraction of \monolayer\ in the powder X-ray diffraction spectra of our
samples does not exceed a few percent. This is possible if during
the annealing process the substitution of the oxygen atoms in the minor
\monolayer\ part for $^{17}$O is easier than the substitution in the
\bilayer\ part of the sample. The spectra
positions and longitudinal relaxation rates $1/T_1$ of \monolayer\ are
known\cite{214Mukuda17O}, and hence, line crossing between \bilayer\ and
\monolayer\ can be avoided when measuring Knight shifts or $1/T_1$ of \bilayer.

\subsection{Quadrupole splitting}\label{sec:efg}
The EFG at the oxygen site is the sum of the fields created by
external ions and
by on-site unclosed $2p$ shells. The former is largely enhanced over the
value obtained by point-charge calculations due to polarization of the
on-site closed shells.
We employ the Sternheimer equation\cite{Sternheimer,EFGHighTc} to describe the
EFG frequencies $^{17}\!\nu^{i}$:
\begin{equation}\label{eq:nu}
^{17}\!\nu^{i} = (1-\gamma_{\infty}){^{17}\!\nu^{i}_{\rm lat}} + (1-R){^{17}\!\nu^{i}_{\rm hole}},
\end{equation}
where $^{17}\!\nu^{i}_{\rm lat}$ is obtained from the EFG frequency of the
point-charge calculation, and $^{17}\!\nu^{i}_{\rm hole}$ is that of the holes in the
$2p$
orbitals. The two Sternheimer parameters $\gamma_{\infty}$ and $R$ are the
antishielding and the shielding factors, respectively. $R$
is approximately 0.1 for oxygen $2p$,\cite{Sternheimer,EFGHighTc} and
we adopt this value. It is difficult to estimate the value of $\gamma_{\infty}$
for O$^{2-}$.
$^{17}$O-NMR experiments in cuprates yielded $\gamma_{\infty}$ of $-8$
(Ref.~\onlinecite{ThurberSpinLadder}) or $-9$ (Ref.~\onlinecite{YBCOTakigawa}).
However, theoretical calculations suggested a wide variety of values between
$-9$ and $-33$,\cite{EFGionic} indicating
that the value is highly dependent on the ionic radius or
covalency.\cite{EFGHighTc}

\begin{table*}
\caption{\label{tab:eqq}Electric field gradients' (EFG) frequencies of
the ruthenates determined from the observed electric quadrupole splittings,
compared with those obtained from point-charge model calculations where the
charges of the oxygen and the ruthenium are -2 and +4, respectively. The signs
 have been adjusted to obtain positive EFG
frequencies along the principal axes. Order-of-magnitude differences
between experimentally determined EFGs and calculated EFGs 
mostly arise from an antishielding effect $(1-\gamma_{\infty})$. Lattice
parameters used for the calculations are based on
Refs.~\onlinecite{Shaked327},~\onlinecite{Chmaissem214}, and \onlinecite{Jones113} for
\bilayer, \monolayer, and \inflayer, respectively.}
\begin{ruledtabular}
\begin{tabular}{ccccccccc}
& $^{17}\!\nu^{\parallel,a}$ & $^{17}\!\nu^{\perp,b}$ & $^{17}\!\nu^\text{c}$ & $\eta$&
$^{17}\!\nu^{\parallel,a}_{\rm lat}$ & $^{17}\!\nu^{\perp,b}_{\rm lat}$ &
$^{17}\!\nu^\text{c}_{\rm lat}$ & $\eta_{\rm lat}$ \\ & MHz & MHz & MHz & & MHz &
MHz & MHz &\\
 \hline
\multicolumn{9}{c}{\bilayer}\\
O(1) & $\mp$0.47 & $\mp$0.47 & $\pm$0.94 & 0 &
-0.0830& -0.0830& 0.1660& 0\\ 
O(2) & $\mp$0.33 & $\mp$0.33 &$\pm$0.66 & 0 &
-0.0456& -0.0456& 0.0910& 0\\
O(3) & $\pm$0.78 &$\mp$0.44 & $\mp$0.34 & 0.11 &
0.1951& -0.0971& -0.0980& 0.004\\ 
 \hline
\multicolumn{9}{c}{\monolayer}\\
O(1) & $\pm$0.77\footnote{H.~Murakawa \textit{et al}., single crystal with an accurate
 sample alignment in Ref.~\onlinecite{214Murakawa17ONMR}. Values are the same in
 Ref.~\onlinecite{214Mukuda17O} within errors.}& $\mp$0.45\footnotemark[1]&
 $\mp$0.32\footnotemark[1] & 0.17\footnotemark[1]& 0.2062 & -0.0986 & -0.1064 &
0.04\\
O(2) & $\mp$0.305\footnotemark[1]& $\mp$0.305\footnotemark[1]& $\pm$0.610\footnotemark[1]&
 0\footnotemark[1] & -0.0437 & -0.0437 & 0.0874 & 0\\
 \hline
\multicolumn{9}{c}{Paramagnetic \inflayer}\\
O & $\pm$0.95\footnote{K.~Yoshimura \textit{et al.}, powder sample in
Ref.~\onlinecite{113NMR}.} & & & & 0.1782 & -0.0891 & -0.0891 & 0
\end{tabular}
\end{ruledtabular}
\end{table*}
Table~\ref{tab:eqq} shows the measured EFG frequencies $^{17}\!\nu^{i}$
determined from the quadrupole splitting of the satellite lines and 
$^{17}\!\nu^{i}_{\rm lat}$ from calculations based on the
point-charge model without considering holes at the oxygen sites. In the calculation,
ions within at least three lattice units have to be included
to obtain sufficient convergence. The spectral EFG frequencies do not feature
any significant
$H$ and $T$ ($<100$~K) dependence, suggesting
that the electronic band structure of \bilayer\ is not strongly modified
by the first-order metamagnetic transition. In fact, de~Haas-van~Alphen\cite{BorzidHvA} and Shubnikov-de~Haas\cite{PerrySdH}
measurements revealed that the changes in the oscillating
frequencies are quite small, supporting a field-induced Stoner transition model
of the itinerant metamagnetism of \bilayer.

In the metallic ruthenate, the Fermi level is located between the
bonding and the antibonding molecular orbitals,\cite{Oguchi214} and hence,
the atomic $2p$ orbitals acquire holes. In
terms of the hole occupancy
$h_{2p}$ of the $2p$ orbital, the induced EFG $^{17}\!\nu^{i}_{\rm hole}$ can be
written as:
\begin{equation}
\begin{pmatrix}
^{17}\!\nu^{x}_{\rm hole}\\
^{17}\!\nu^{y}_{\rm hole}\\
^{17}\!\nu^{z}_{\rm hole}
\end{pmatrix}
 = \frac{2}{5}eh_{2p}\left<r^{-3}\right>_{2p}
\begin{pmatrix}
-1\\ -1\\ 2 \end{pmatrix},
\end{equation}
for the $2p_z$ orbital, for instance. Taking $\left<r^{-3}\right>_{2p}$ to be
3.63~a.u., which is a 70\% value of that for a free atom,\cite{YBCOTakigawa,
17OESR}
$^{17}\!\nu^{i}_{\rm hole} = h_{2p}(-1.33, -1.33, 2.66)$~MHz. Thus, the total EFG frequency
at each site is

(For the apical sites)
\begin{equation}
\begin{pmatrix}
^{17}\!\nu^{a,b}\\ ^{17}\!\nu^{c}
\end{pmatrix}
= (1-\gamma_{\infty}){^{17}\!\nu^{i}_{\rm lat}} + (1-R)h_{2p_{a,b}}\begin{pmatrix}
1.33~\text{MHz}\\ -2.66~\text{MHz} \end{pmatrix},
\end{equation}

(or for the in-plane sites)
\begin{align}
\begin{pmatrix}
^{17}\!\nu^{\parallel}\\ ^{17}\!\nu^{\perp}\\ ^{17}\!\nu^{c}
\end{pmatrix}
= (1-\gamma_{\infty}){^{17}\!\nu^{i}_{\rm lat}} + (1-R)h_{2p_{c}}\begin{pmatrix}
-1.33~\text{MHz}\\ -1.33~\text{MHz}\\ 2.66~\text{MHz} \end{pmatrix}\nonumber\\
+ (1-R)h_{2p_{a,b}}\begin{pmatrix} -1.33~\text{MHz}\\ 2.66~\text{MHz}\\
-1.33~\text{MHz} \end{pmatrix}.
\end{align}
The antishielding factor $\gamma_{\infty}$ cannot be solved from the above
equations because of the dipole-field requirement $\sum_i {^{17}\!\nu^i} = 0$.
\begin{table}
\caption{\label{tab:h2p} Hole occupancies in the oxygen $2p$ orbitals obtained
from
the EFG analysis. Although the derivation of a reliable value of the
Sternheimer antishielding
factor $\gamma_{\infty}$ is technically difficult, we argue that
$\gamma_{\infty}=-9$ yields physically reasonable results (see text).
$h_{2p}^{\rm total}$ denotes the total hole number per
ruthenium atom. The given literature values of band-structure
calculations have been estimated by integrating the DOS above
$E_{\rm F}$.}
\begin{ruledtabular}
\begin{tabular}{lcccr}
& \multicolumn{3}{c}{NMR EFG Analysis}& Band Calc.\\
& \multicolumn{3}{c}{$\gamma_{\infty}$}&\\
& -8& -9& -10&\\
\hline
\multicolumn{5}{c}{\bilayer}\\
$h_{2p_{a,b}}^{\rm O(1)}$& 0.056& 0.075& 0.092&\\
$h_{2p_{a,b}}^{\rm O(2)}$& 0.005& 0.009& 0.021&\\
$h_{2p_{a,b}}^{\rm O(3)}$& 0.178& 0.199& 0.218&\\
$h_{2p_{c}}^{\rm O(3)}$& 0.232& 0.256& 0.277&\\
$h_{2p}^{\rm O(3)}$& 0.410& 0.455& 0.495&
 0.28\footnote{After Fig.~4 in Ref.~\onlinecite{Hase327}.}\\
$h_{2p}^{\rm total}$& 0.866& 1.00& 1.12&
 0.76\footnotemark[1]\\
\hline
\multicolumn{5}{c}{\monolayer}\\
$h_{2p_{a,b}}^{\rm O(1)}$& 0.183& 0.202& 0.219&\\
$h_{2p_{c}}^{\rm O(1)}$& 0.276& 0.301& 0.323&\\
$h_{2p}^{\rm O(1)}$& 0.459& 0.503& 0.542&
 0.27\footnote{After Fig.~3 in Ref.~\onlinecite{Oguchi214}.}
 0.28\footnote{After Fig.~4 in Ref.~\onlinecite{Hase214}.}\\
$h_{2p_{a,b}}^{\rm O(2)}$& -0.001& 0.012& 0.023&\\
$h_{2p}^{\rm O(2)}$& -0.002& 0.024& 0.046&
 0.15\footnotemark[2]
 0.09\footnotemark[3]\\
$h_{2p}^{\rm total}$& 0.915& 1.05& 1.18&
 0.84\footnotemark[2]
 0.75\footnotemark[3]\\
\hline
\multicolumn{5}{c}{Paramagnetic \inflayer}\\
$h_{2p}^{\rm O}$& 0.120& 0.143& 0.164&\\
$h_{2p}^{\rm total}$& 0.718& 0.859& 0.984&\\
\end{tabular}
\end{ruledtabular}
\end{table}
Table~\ref{tab:h2p}
shows the hole occupancy in each orbital estimated from various values for 
$\gamma_{\infty}$ and
from band-structure calculations in literature. In this analysis, 
$\gamma_{\infty} < -8$ is necessary to make all densities positive. 
In the following we adopt the same value used in the case of the cuprates 
$\gamma_{\infty} =
-9$\cite{YBCOTakigawa,EFGHighTc}. This leads to a
physically reasonable set of values for $h_{2p}$.

It should be noted that
most of the holes reside at the in-plane sites (O(3) for \bilayer\ and O(1) for
\monolayer), reflecting their quasi-two dimensionality. The large magnitude of
the in-plane hole occupancy is in agreement with the strongly anisotropic
conductivity
($\rho_{c}/\rho_{ab} \simeq 300$ for \bilayer\cite{SIkeda} and $\agt 400$ fo
\monolayer\cite{214review}) and with spin densities derived from the Knight
shift measurements, which are discussed later.

Another remark concerns the relationship between hole occupancy and bonding
covalency. 
In general, a hole in an antibonding molecular orbital introduces covalent
character in the
original ionic bond. Since a covalent bonding is sensitive to the Ru-O-Ru angle,
the covalency can be related to a structural instability of the RuO$_6$
octahedra. In \monolayer,
Oguchi argued that the high $h_{2p_c}$ in the $4d_{zx,yz}$-$2p_{c}$ (one of
$p_{\pi}$) antibonding orbitals may account for an absence of a rotation of the 
RuO$_6$ octahedra.\cite{Oguchi214} Indeed, in Table~\ref{tab:h2p}, 
$h_{2p_{c}}$,
which is hybridized with $d_{zx,yz}$, has a larger value ($=0.301$) than $h_{2p_{a,b}}$
($=0.202$) hybridizing with $d_{xy}$ at the in-plane sites. In addition, a 
smaller value
of $h_{2p_{c}}$ ($=0.256$) at the in-plane site is obtained for \bilayer, where
the RuO$_6$ octahedra are rotated by 7\degree.\cite{Shaked327} We consider that
the existence of the rotation in \bilayer\ is related to a weaker covalency
in the $d_{zx,yz}$-$2p_{c}$ bond than that in \monolayer. For comparison with 
\inflayer, which has an octahedra tilt, the hole occupancy per ruthenium
atom
$h_{2p}^{\rm total}$ can be used. \inflayer\ has the smallest value of $h_{2p}^{\rm total}$
in Table~\ref{tab:h2p}, which means the weakest covalency per ruthenium atom. This
might be related to the occurrence of the tilt in \inflayer\ as well as the
rotation in \bilayer.

We point out experimentally that the hole content of the antibonding orbitals
may be substantially relevant to the distortion of the RuO$_6$
octahedra even in Sr$_{n+1}$Ru$_n$O$_{3n+1}$ which have the
same tolerance factor.

\subsection{Knight shift}\label{sec:shift}
Generally, the observed Knight shift $^{17}\!K_{\rm obs}$ (or the uniform
susceptibility $\chi_{\rm bulk}$) for a given field direction $i\,(=x,y,z)$ is
the sum of the temperature
independent orbital part $^{17}\!K_{\rm orb}^i$ ($\chi_{\rm bulk, orb}^i$), the
diamagnetic
part $^{17}\!K_{\rm dia}^i$ ($\chi_{\rm bulk, dia}$), and the temperature
dependent spin
part $^{17}\!K^i_{\rm s}$ ($\chi^i_{\rm bulk, s}$), and are expressed as
follows:
\begin{align}\label{eq:knightshift}
^{17}\!K_{\rm obs}^{i}(T)&= {^{17}\!K_{\rm orb}^{i}} + {^{17}\!K_{\rm dia}^{i}} +
{^{17}\!K_{\rm s}^{i}(T)},\\
\chi_{\rm bulk}^{i}(T) &= \chi_{\rm bulk, orb}^{i} + \chi_{\rm bulk, dia} +
\chi_{\rm bulk, s}^{i}(T),\\
^{17}\!K_{\rm s}^{i}(T)&=\frac{^{17}\!A_{\rm s}^{i}}{N_{\rm A}\mu_{\rm B}}\chi_{\rm
s}^{i}(T).\label{eq:knightshift3}
\end{align}
Here, $N_{\rm A}$, $\mu_{\rm B}$, and $^{17}\!A_{\rm s}^{i}$ denote Avogadro's
number, the Bohr magneton, and the spin hyperfine coupling constant,
respectively. 
Note that the spin part of $M/H_{\rm res}$ must be used instead of
the differential susceptibility $\chi_{\rm bulk, s}$ in eq.~\eqref{eq:knightshift3}
unless $M$ is proportional to $H$.

\begin{figure}[tb]
\centering
\includegraphics[width=\linewidth]{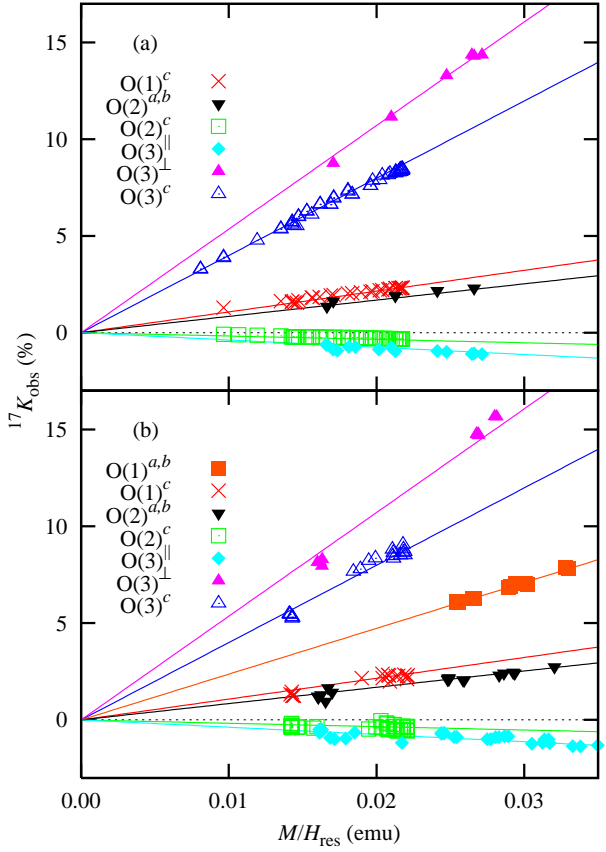}
\caption{\label{fig:kchiplot}
(color online). $^{17}\!K_{\rm obs}$ vs $M/H_{\rm res}$ plots for \bilayer. The
fitted hyperfine
coupling constant for each curve is listed in Table~\ref{tab:hyperfine}.
(a) Temperature
varied data points between 1.7~K and 100~K at low field $\simeq 4$~T and
(b) field varied points between 3~T and 13~T
at 1.7~K. Below 7~T $M$ is
measured with a commercial SQUID magnetometer. For higher fields, $M$ is
interpolated (7~T -- 8~T), derived from 2.8~K data given in 
Ref.~\onlinecite{PerryPRL} (8~T -- 11~T), and extrapolated ($> 11$~T).} 
\end{figure}
Figure~\ref{fig:kchiplot} shows a $^{17}\!K_{\rm obs}$ vs $M/H_{\rm res}$ plot of
\bilayer.
$^{17}\!K_{\rm obs}$ and $M/H_{\rm res}$ are measured at
$\sim 4$~T by changing temperature 
in a range of 1.7-100~K passing through the susceptibility maximum at 16~K
(Fig.~\ref{fig:kchiplot}(a)) and at 1.7~K by changing the magnetic field in a
range
of 3-13~T across the metamagnetic field (Fig.~\ref{fig:kchiplot}(b)). For the
[100] direction, the
temperature dependence for the O(1) site could not be measured due to the
crossing of the lines to other sites. Linear relations are obtained
in both cases and the solid lines are linear fits. The slopes of the
$K$ vs $M/H_{\rm res}$ plots for those two sets of data give the same hyperfine
coupling constant $^{17}\!A_{\rm s}^i$. The
orbital component of the Knight shifts are negligibly small since the lines intersect
the origin. The obtained spin hyperfine coupling constants $^{17}\!A_{\rm s}^{i}$
are listed in Table~\ref{tab:hyperfine}.

We extended the previous tight-binding approach\cite{214Imai} by including the
magnetic dipolar fields from neighboring sites: 

\begin{equation}\label{eq:knightshift-tb}
^{17}\!K_{\rm obs}^{i}={^{17}\!K_{\rm off-site}^{i}} + {^{17}\!K_{\rm on-site}^{i}}.
\end{equation}
The first term denotes the shift of the field produced by the moment of the adjacent
sites, mostly of the
nearest-neighbor ruthenium sites, and the second term is produced by
the transferred spin densities in the oxygen on-site orbitals. In our case, the
latter plays a major role as a result of a large local spin density.

 The off-site part at the oxygen $n$ site is written in terms of local
 susceptibilities at neighboring $m$ sites $\chi^i_m$:
\begin{align}\label{eq:koffsite}
^{17}\!K_{\rm off-site, orb}^{i}&=\frac{1}{N_{\rm A}\mu_{\rm B}}
\sum_{m(\neq n)}\frac{\partial^2}{\partial
i^2}\left(\frac{1}{r_{mn}}\right)\chi^i_{m,\rm orb},\nonumber\\
^{17}\!K_{\rm off-site, dia}^{i}&=\frac{1}{N_{\rm A}\mu_{\rm B}}
\sum_{m(\neq n)}\frac{\partial^2}{\partial i^2}\left(\frac{1}{r_{mn}}\right)\chi_{m,\rm dia},\nonumber\\
^{17}\!K_{\rm off-site, s}^{i}(T)&=\frac{1}{N_{\rm A}\mu_{\rm B}}
\sum_{m(\neq n)}\frac{\partial^2}{\partial i^2}\left(\frac{1}{r_{mn}}\right)\chi^i_{m,\rm s}(T).
\end{align}
Here, $r_{mn}$ is the distance between two sites $m$ and $n$.
The summations are taken over a distance of four lattice units to ensure
convergence. Since $\chi_{\rm bulk, s}$ of \bilayer\ and
\inflayer\ are strongly enhanced
($>1\times 10^{-2}$~emu/Ru-mol), the orbital part or diamagnetic part of $\chi$
(typically $10^{-4}$~emu/Ru-mol\cite{214Imai}) for those compounds can be
neglected. For \monolayer, the estimated values of
${^{17}\!K_{\rm
orb}} + {^{17}\!K_{\rm dia}}$ are also given in Table~\ref{tab:hyperfine}. For all
three compounds, the off-site
hyperfine coupling constants from the ruthenium sites to the oxygen sites
 $\lvert\sum_{m={\rm Ru}}\frac{\partial^2}{\partial i^2}(\frac{1}{r_{mn}})\rvert$ is small ($<5$~kOe/$\mu_{\rm B}$).
Therefore, the $^{17}$O Knight shifts are dominated by the on-site spin parts,
which we will discuss next.

\begin{table}[tb]
\caption{\label{tab:hyperfine} $^{17}$O Hyperfine coupling constants $^{17}\!A_{\rm
s}$. $^{17}\!A_{\rm s}$ of \bilayer\ are obtained from fittings of $^{17}\!K_{\rm obs
}$ vs $M/H_{\rm res}$ plots
(Fig.~\ref{fig:kchiplot}). Estimated non-spin parts of the Knight shifts
${^{17}\!K_{\rm
orb}}+ {^{17}\!K_{\rm dia}}$ for \monolayer\ are also shown. For the other
compounds, the non-spin parts are negligibly small.}
\begin{ruledtabular}
\begin{tabular}{lcc}
& $^{17}\!A_{\rm s}$& ${^{17}\!K_{\rm orb}} + {^{17}\!K_{\rm dia}}$ \\
& (kOe/$\mu_{\rm B}$)& (\%)\\
\hline
\multicolumn{3}{c}{\bilayer}\\
O(1)$^{a,b}$& $13.2\pm0.2$& \\
O(1)$^{c}$& $6.0\pm0.2$& \\
O(2)$^{a,b}$& $4.7\pm0.3$& \\
O(2)$^{c}$& $-0.97\pm0.05$& \\
O(3)$^{\parallel}$& $-2.1\pm0.3$& \\
O(3)$^{\perp}$& $29.9\pm0.3$& \\
O(3)$^{c}$& $22.3\pm0.4$& \\
\hline
\multicolumn{3}{c}{\monolayer}\\
O(1)$^{\parallel}$& $-7.9\pm1$\footnote{H.~Murakawa \textit{et al}., single crystal with an accurate
 sample alignment in Ref.~\onlinecite{214Murakawa17ONMR}. Knight shifts are taken
 below 4~K and are the same in Ref.~\onlinecite{214Imai} within errors.}&
 0.013\\ O(1)$^{\perp}$& $33.1\pm1$\footnotemark[1]& -0.006\\
O(1)$^{c}$& $23.7\pm1$\footnotemark[1]& -0.005\\
O(2)$^{a,b}$& $6.4\pm1$\footnotemark[1]& -0.003\\
O(2)$^{c}$& $1.9\pm1$\footnotemark[1]& 0.004\\
\hline
\multicolumn{3}{c}{Paramagnetic \inflayer}\\
O$^{\parallel}$& $6.3\pm2$\footnote{Anisotropic Knight shifts are estimated
from the powder patterns
 above $T_{\rm C}$ by pattern simulations. The spectra used in the study of
 Ref.~\onlinecite{113NMR} were provided by Yoshimura {\it et al.}}& \\
 O$^{\perp}$& $21\pm2$\footnotemark[2]& 
\end{tabular}
\end{ruledtabular}
\end{table}
 Assuming that only the ruthenium orbitals contribute to the orbital part of the
 bulk susceptibility, the on-site Knight shifts at the oxygen sites arise
 solely from the spin part, 
\begin{equation}\label{eq:konsite}
^{17}\!K_{\rm on-site}^{i}(T)={\sum_{k=\{2p_{a,b,c}, 2s\}}} 
\frac{^{17}\!A_{k}^{i}}{N_{\rm A}\mu_{\rm B}}\chi_{k, {\rm s}}^{i}(T),
\end{equation}
where $^{17}\!A_k^{i}$ is the hyperfine coupling constant
and $\chi_{k, {\rm s}}^{i}(T)$ is the transferred spin density, and the subscript 
$k\,(=2p,2s)$ specifies the orbital. Since the isotropic
hyperfine coupling of the $2s$ orbital $^{17}\!A_{2s}$ is quite large due to its
Fermi contact to the nucleus ($^{17}\!A_{2s} \sim 3000$~kOe/$\mu_{\rm
B}$\cite{YBCOTakigawaNATO}), isotropic shifts still dominate even with
a tiny spin density in the $2s$ orbitals. The hyperfine couplings of $2p$ orbitals
have an anisotropic dipole character, $2\,{^{17}\!A_{2p}}(-1, -1, 2)$ for each
$2p$ orbital, where
${^{17}\!A_{2p}} = \frac{2}{5}\mu_{\rm B}\left<r^{-3}\right>_{2p} \simeq
91$~kOe/$\mu_{\rm B}$.\cite{17OESR,YBCOTakigawa}

Next, we assume that the ratios of transferred orbital spin densities 
$\chi_{k, s}/\chi_{\rm bulk, s}\,(k=2p,2s)$ are independent of the field
orientation even in the case of anisotropic spin susceptibilities. Non-bonding
$2p_{\sigma}$ orbitals can be ignored because states arising from those orbitals
 are much below $E_{\rm F}$. Thus, the total on-site hyperfine
coupling constant $^{17}\!A_{\rm s, on-site}$ for each field orientation is
simplified to:

(For the apical sites)
\begin{equation}\label{eq:konsite-apical}
\begin{pmatrix}
^{17}\!A_{\rm s,on-site}^{a,b}\\
^{17}\!A_{\rm s,on-site}^{c}
\end{pmatrix}
=
\begin{pmatrix}
1& 1\\
-2& 1
\end{pmatrix}
\begin{pmatrix}
2\,{^{17}\!A_{2p}}\chi_{2p_{x,y}, {\rm s}}/\chi_{\rm bulk, s}\\
 {^{17}\!A_{2s}}\chi_{2s, {\rm s}}/\chi_{\rm bulk, s}
\end{pmatrix},
\end{equation}

(or for the in-plane sites)
\begin{align}\label{eq:konsite-inplane}
\begin{pmatrix}
^{17}\!A_{\rm s,on-site}^{\parallel}\\ 
^{17}\!A_{\rm s,on-site}^{\perp}\\
^{17}\!A_{\rm s,on-site}^{c}
\end{pmatrix}
&=\nonumber\\
\begin{pmatrix}
-1& -1& 1\\
2& -1& 1\\
-1& 2& 1
\end{pmatrix}
&\begin{pmatrix}
2\,{^{17}\!A_{2p}}\chi_{2p_{x,y}, {\rm s}}/\chi_{\rm bulk, s}\\
2\,{^{17}\!A_{2p}}\chi_{2p_{c}, {\rm s}}/\chi_{\rm bulk, s}\\
{^{17}\!A_{2s}}\chi_{2s, {\rm s}}/\chi_{\rm bulk, s}.
\end{pmatrix},
\end{align}

Since the spin density at strontium sites can be ignored, the bulk spin
susceptibility $\chi_{\rm bulk, s}$ is the sum of the spin densities in the oxygen and
the ruthenium orbitals,
\begin{equation}
\chi_{\rm bulk, s} = \chi^{\rm Ru}_{\rm s} + \chi^{\rm total}_{2p, {\rm s}} + \chi^{\rm
total}_{2s, {\rm s}}.
\end{equation}
For \bilayer, the total spin densities over all oxygen sites per ruthenium
atom $\chi_{2p, {\rm s}}^{\rm total}$ and $\chi_{2s, {\rm s}}^{\rm total}$ are:
\begin{align}
\chi_{2p, {\rm s}}^{\rm total} &= \chi_{2p_{a,b}, {\rm s}}^{\rm O(1)} +2\chi_{2p_{a,b}, {\rm s}}^{\rm O(2)}
+ 2\chi_{2p_{a,b}, {\rm s}}^{\rm O(3)} + 2\chi_{2p_{c}, {\rm s}}^{\rm O(3)},\\
\label{eq:chitotals}
\chi_{2s, {\rm s}}^{\rm total} &= \frac{1}{2}\chi_{2s, {\rm s}}^{\rm
O(1)} + \chi_{2s, {\rm s}}^{\rm O(2)} + 2\chi_{2s, {\rm s}}^{\rm O(3)}.
\end{align}
Thus, the observed anisotropic Knight shifts can be decomposed into the local
spin
density in each orbital from eqs.~\eqref{eq:knightshift}-\eqref{eq:chitotals}.

\begin{table}[tb]
\caption{\label{tab:spindensity}Orbital spin densities obtained from Knight-shift measurements and the density of states at the Fermi level $D(E_{\rm F})$
at the oxygen site
extracted from the results of band-structure calculations. Spin densities and
$D(E_{\rm F})$ are normalized by the bulk spin susceptibility per ruthenium atom
$\chi_{\rm bulk, s}$.}
\begin{ruledtabular}
\begin{tabular}{lccc}
& $\chi({\rm per~orb.})/\chi_{\rm bulk, s}$& $\chi({\rm per~Ru})/\chi_{\rm bulk, s}$&
$D(E_{\rm F})$\\
 & \%& \%& \% \\
\hline
\multicolumn{4}{c}{\bilayer}\\
$\chi_{2p_{a,b}, {\rm s}}^{\rm O(1)}$& $1.71\pm0.06$& $1.71\pm0.06$& 3\footnote{Reference~\onlinecite{Singh327}.}\\
$\chi_{2s, {\rm s}}^{\rm O(1)}$& $0.360\pm0.005$& $0.180\pm0.003$\\
$\chi_{2p_{a,b}, {\rm s}}^{\rm O(2)}$& $1.50\pm0.06$& $3.00\pm0.11$& 2\footnotemark[1]\\
$\chi_{2s, {\rm s}}^{\rm O(2)}$& $0.093\pm0.007$& $0.093\pm0.007$\\
$\chi_{2p_{a,b}, {\rm s}}^{\rm O(3)}$& $6.38\pm0.08$& $12.8\pm0.16$\\
$\chi_{2p_{c}, {\rm s}}^{\rm O(3)}$& $5.03\pm0.09$& $10.1\pm0.18$\\
$\chi_{2p, {\rm s}}^{\rm O(3)}$& & $22.9\pm0.3$&
 20\footnotemark[1]
 21\footnote{FLAPW within LSDA. Taken from figures in Ref.~\onlinecite{Hase327}.}\\
$\chi_{2s, {\rm s}}^{\rm O(3)}$& $0.557\pm0.007$& $1.113\pm0.013$\\
$\chi_{2p, {\rm s}}^{\rm total}$& & $27.5\pm0.3$& 25\footnotemark[1] 25\footnotemark[2]\\
$\chi_{2s, {\rm s}}^{\rm total}$& & $1.39\pm0.015$\\
\hline
\multicolumn{4}{c}{\monolayer}\\
$\chi_{2p_{a,b}, {\rm s}}^{\rm O(1)}$& $8.4\pm0.24$& $16.7\pm0.5$ \\
$\chi_{2p_{c}, {\rm s}}^{\rm O(1)}$& $6.8\pm0.23$& $13.6\pm0.5$\\
$\chi_{2p, {\rm s}}^{\rm O(1)}$& & $30.3\pm0.8$&
15\footnote{LAPW within LDA. Taken from Fig.~3 in Ref.~\onlinecite{Oguchi214}.}
26\footnote{LAPW within LDA. Taken from Fig.~2 in Ref.~\onlinecite{Singh214}.}
21\footnote{FLAPW within LSDA. Taken from figs in Ref.~\onlinecite{Hase214}.}\\
$\chi_{2s, {\rm s}}^{\rm O(1)}$& $0.54\pm0.02$& $1.11\pm0.04$\\
$\chi_{2p_{a,b}, {\rm s}}^{\rm O(2)}$& $1.2\pm0.24$& $5.0\pm0.9$&
3\footnotemark[3] 4\footnotemark[4] 4\footnotemark[5]\\
$\chi_{2s, {\rm s}}^{\rm O(2)}$& $0.16\pm0.024$& $0.33\pm0.05$\\
$\chi_{2p, {\rm s}}^{\rm total}$& & $35.3\pm1.2$& 
18\footnotemark[3] 29\footnotemark[4] 25\footnotemark[5]\\
$\chi_{2s, {\rm s}}^{\rm total}$& & $1.41\pm0.06$\\
\hline
\multicolumn{4}{c}{Paramagnetic \inflayer}\\
$\chi_{2p, {\rm s}}^{\rm O}$& $3.6\pm0.5$& $22\pm3$& 
21\footnote{LMTO within LSDA in Ref.~\onlinecite{Allen113}.}
30\footnote{Approximate value in Ref.~\onlinecite{Mazin113}.}\\
$\chi_{2s, {\rm s}}^{\rm O}$& $0.54\pm0.05$& $1.6\pm0.15$&
\end{tabular}
\end{ruledtabular}
\end{table}

 Table~\ref{tab:spindensity} displays the local spin densities in the oxygen
 orbitals obtained from Knight shifts and the DOS at
 $E_{\rm F}$, $D(E_{\rm F})$ from band-structure calculations in literature.
The spin density is generally enhanced over the value of $D(E_{\rm F})$ which
comes out of band-structure calculations due to
Stoner enhancement and untreated electron-electron correlations. The bulk
susceptibility of \bilayer\ is the product of the
value inferred from the electronic specific heat and the large Wilson
ratio 10 (Ref.~\onlinecite{SIkeda}). 
If such a Stoner enhancement varies between bands and/or site-by-site, the
enhancement factors applied to local spin
densities would differ from orbital to orbital.
 In Table~\ref{tab:spindensity}, the normalized values of spin densities and
of $D(E_{\rm F})$ are essentially comparable for each oxygen site. The Stoner
enhancement of these ruthenates appears to be independent of the
orbitals.

According to this analysis, the total spin density at the oxygen sites amounts to
20-40\%
of the bulk susceptibility. This is quite large for a metallic oxide.
Indeed, a polarized neutron scattering
study on the exchange enhanced paramagnet Ca$_{1.5}$Sr$_{0.5}$RuO$_4$ by 
A.~Gukasov and co-workers (Ref.~\onlinecite{PNSCSRO4}) revealed
that the in-plane site O(1) and the apical site O(2) have 30\% and 4\% of
the total moment, respectively, and the totaling one third of moments resides
at the oxygen
sites. In the case of
\monolayer\ the values are similar to our results shown in
Table~\ref{tab:spindensity}.

It is found that higher dimensional members have a much smaller fraction of spin
densities
at the oxygen sites $\chi_{2p, {\rm s}}^{\rm total}/\chi_{\rm bulk, s}$, while
the variation of the hole occupancies
$h_{2p}^{\rm total}$ in Table~\ref{tab:h2p} is small. Therefore, we can deduce that
\inflayer\ has a largely enhanced $D(E_{\rm F})$ especially at the ruthenium
site because of the on-site Coulomb
interactions at the ruthenium site. In a Stoner mechanism, the partially localized
character of the $4d$ electrons and resulting narrow bands and high $D(E_{\rm
F})$ are favorable for itinerant ferromagnetism and metamagnetism, as is
 systematically observed in $3d$
and $4d$ pure metals. The ferromagnetism of \inflayer\ could be related to the
smallness of the transferred spin density at the oxygen site.

In addition to the ferromagnetic fluctuations in \bilayer, incommensurate antiferromagnetic fluctuations
were found,\cite{KitagawaPRL,CapognaINS} which are similar to those in
\monolayer.\cite{214SidisINS} These incommensurate antiferromagnetic
fluctuations arise from the Fermi-surface nesting of one-dimensional
Fermi-surface sheets,
when the small spin density at the apical oxygen site weakens the transfer
between
RuO$_2$ layers, resulting in the low-dimensional character. We thus clarified
the important roles of hole occupation and spin density in different oxygen
orbitals and their relevance for the ferromagnetic and antiferromagnetic character
of the ruthenates Sr$_{n+1}$Ru$_n$O$_{3n+1}$.

\section{Conclusion}
 In conclusion, we have investigated microscopic hole occupancies and transferred
 spin densities at the oxygen sites by analyzing the $^{17}$O-NMR electric
 field gradients and anisotropic Knight shifts of Sr$_{n+1}$Ru$_n$O$_{3n+1}$ ($n=1,2,\infty$).
 The hole occupancies at the oxygen sites are
 quite large, which is unexpected for a non-doped oxide. This gives evidence that
 the  
 covalent character of the Ru-O bonding plays an important role in these compounds.
 The 
 tilting (\inflayer) and the rotation (\bilayer) of the RuO$_6$ octahedra
 may account for a weaker
 covalency, which originates from insufficient holes in the
 $p_{\pi}$ antibonding orbitals which is not the case in the undistorted 
 \monolayer.
 
 The total spin densities at the oxygen sites derived from anisotropic
 $^{17}$O-Knight shifts are also high (20-40\% of the bulk spin
 susceptibility), which is, however, in good agreement with  
 band-structure
 calculations and results obtained from polarized neutron experiments 
 reported in literature. Although the total oxygen hole occupancies per ruthenium atom do
 not significantly change among the series Sr$_{n+1}$Ru$_n$O$_{3n+1}$, the spin
 densities at the oxygen sites decrease with increasing $n$. This
 suggests that higher dimensional members of the series have narrower bands
 and, as a result, metamagnetism and itinerant ferromagnetism are induced in 
 \bilayer\ and \inflayer, respectively. It is shown that the enhancement of the
 spin density at the oxygen site is comparable with that at the ruthenium site.
 The spin density at the apical oxygen site (O(2) in \monolayer, O(1) and O(2) in
 \bilayer) is much smaller than that at the in-plane oxygen site, which is
 responsible for the
 strong two-dimensional character in those compounds. The low
 dimensionality might lead to the occurrence of one-dimensional Fermi-surface
 nesting. This would explain the incommensurate antiferromagnetic fluctuations, which emerge in \monolayer\ and in \bilayer.
 Structural distortions and magnetism in Sr$_{n+1}$Ru$_n$O$_{3n+1}$ are
 strongly related with the covalent character of
 the Ru~$4d$-O~$2p$ molecular orbitals and the spin densities at the oxygen sites.

\begin{acknowledgments}
We thank H.~Yaguchi, M.~Kriener, E.~M.~Forgan, and H.~Ikeda for stimulating
discussions and comments. This work was in
part supported by the Grants-in-Aid for Scientific Research from JSPS and MEXT
of Japan and by the 21COE program ``Center for Diversity and
Universality in Physics'' from MEXT of Japan.
\end{acknowledgments}


\bibliography{resubmit} 

\end{document}